\documentclass[twocolumn,amsmath,amssymb, pra]{revtex4}

\usepackage{graphicx, subfigure}
\usepackage{bm}

\begin{document}

\title{A proposal for a scalable universal bosonic simulator using individually trapped ions}

\author{Hoi-Kwan Lau\footnote{kero.lau@utoronto.ca} and Daniel F. V. James}
\affiliation{Center for Quantum Information and Quantum Control (CQIQC),
 Department of Physics, University of Toronto, 60 Saint George Street, Toronto, M5S 1A7 Ontario, Canada}

\date{\today}

\begin{abstract}
We describe a possible architecture to implement a universal bosonic simulator (UBS) using trapped ions.  Single ions are confined in individual traps, and their motional states represent the bosonic modes.  Single-mode linear operators, nonlinear phase-shifts, and linear beam splitters can be realized by precisely controlling the  trapping potentials.  
All the processes in a bosonic simulation, except the initialization and the readout, can be conducted beyond the Lamb-Dicke regime.  Aspects of our proposal can also be applied to split adiabatically a pair of ions in a single trap.
\end{abstract}


\maketitle

\section{Introduction}

Coherently manipulated photons have been proposed to be a good candidate for testing the foundation of quantum mechanics \cite{Weihs:1998p12376}, performing quantum computations \cite{Knill:2001p8225, OBrien:2007p12623}, conducting high precision measurements \cite{Nagata:2007p12427}, and other many applications.  
However, because of poor sources, detection inefficiencies, and weak photon-photon interactions, implementing these proposals for large scale devices is very difficult. 
Some other well-controlled quantum system can simulate the photonic system if it also exhibits bosonic properties \cite{Feynman:1982p10908}.  More specifically, a universal bosonic simulator (UBS) should be able to reproduce the evolution of a bosonic system under the most general form of Hamiltonian.
This requirement is not too stringent as the evolution can be approximated to arbitrary accuracy by a sequence of basic operators that belong to a universal set \cite{Lloyd:1996p10335}.  Lloyd and Braunstein \cite{Lloyd:1999p11945} proved that the simplest universal set of basic operators comprised of all single mode linear operators, at least one multi-mode operator, and at least one nonlinear element.  Efficiently performing these basic operations is hence necessary for any implementation of the UBS.

Ion traps are a suitable candidate for implementing a UBS, in which a high degree of controllability has been demonstrated.
\cite{Haffner:2008p3364}.  The motion of laser cooled ions is quantum in nature, and the excitations of the motional states, i.e. phonons, exhibit bosonic behaviour.  The collective displacement and momentum of the ions are analogous to the quadratures of light fields.  Any arbitrary motional state can be created by combining techniques such as sideband transition \cite{Wineland:1998p10591}, parametric amplification \cite{Heinzen:1990p12462}, and adiabatic passage \cite{Cirac:1994p11995}; in particular, the creation of Gaussian states \cite{Meekhof:1996p11999} and non-classical states \cite{Meekhof:1996p11999, Monroe:1996p12537} from the ground state have been experimentally demonstrated.  When applied to non-ground states, some of these techniques can achieve single phonon linear or nonlinear operations.  Interaction between phonon modes at the few-quanta level have also been observed.  For example, nonlinear beam splitting on a single ion has been performed by 
applying a Raman field \cite{Leibfried:2002p11781}; coupling two phonon modes have also been demonstrated through the Coulomb interaction between two separately trapped ions \cite{Brown:2011p11810, Harlander:2011p11811}, or two ions in the same trap \cite{Roos:2008p12727, Nie:2009p12001}. 

The idea of using trapped ions system for UBS was first proposed by Wineland \textit{et al.} \cite{Wineland:1998p11782}. Their simulator is a chain of ions trapped by a harmonic potential.  If all three motional degrees of freedom are incorporated, a trap with $N$ ions can at most offer 3$N$ phonon modes.  This architecture is in principle scalable, because the number of ions in a trap is not fundamentally limited.  However, the addition of ions will narrow the frequency gap between phonon modes; sideband transition should be conducted slowly to avoid significant errors \cite{LauJames:boson}.  In addition, a measurement on one phonon mode via resonance fluorescence may cause significant heating of the ion chain, which distorts the states of other phonon modes.  These shortcomings limit the number of modes and the population of phonons that can be simulated accurately.

The problem of an excess of ions in a single trap also appears in ion trap quantum computing \cite{Haffner:2008p3364}.  Kielpinski, Monroe, and Wineland (KMW) \cite{Kielpinski:2002p7096} have proposed a modified approach in which ion qubits are stored in separate locations of an array of traps, so that the manipulation on one qubit negligibly affects the others.  
Considerable advances in experimental realization of these ideas have been made in the past few years \cite{Reichle:2006p11931, Blakestad:2009p8015}; in particular , entanglement gates have been performed on ions which were initially far-separated, and ions have been moved between traps.

In this paper, we propose using the KMW architecture to implement a UBS on trapped ions system.  We consider each ion to be stored in a separate harmonic trap, in which only one bosonic mode is present.  All single mode operations can be conducted by either changing the storage trap potential or by laser manipulations.  The linear beam splitter is based on the Coulomb interaction, which is the same principle as the kinetic energy exchange in Ref. \cite{Brown:2011p11810, Harlander:2011p11811}.  However, we require the distance between the ions to be variable in order to speed up the process.  Ions can be transported in specific trajectories that do not cause motional excitations \cite{Torrontegui:2011p9810}.  The advantages of our scheme are that the quality of each operation is independent of the number of modes involved in the simulation, and the initialization and readout of any one mode will not distort the others.

We begin by presenting the setup and the physical model of our proposal in Sec. \ref{model}.  Useful mathematical techniques is introduced in Sec. \ref{LR theory}.  The implementation of single mode operations is then introduced in Sec. \ref{single}.  In Sec. \ref{phonon beam splitter}, we show that a linear beam splitter can be implemented by precisely combining and splitting two traps through changing the quadratic and quartic potentials. 
Initialization and readout of phonon states are presented in Sec. \ref{initialization}.  This article is concluded in Sec. \ref{conclusion} with some discussions.

\section{Ions in harmonic traps \label{ion}}

\subsection{Model \label{model}}

We assume ions are tightly trapped in the $y$ and the $z$ directions by a strong ac field while a weaker dc potential is applied along the $x$ direction.  We assume this configuration would effectively restrict the ion to move along only the  $x$ direction because the excitations in other directions are negligible.  

\begin{figure}
\centering
\includegraphics{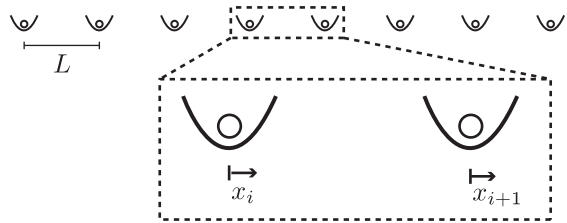}
\caption{\label{fig:config} The configuration of our ion trap UBS is an array of storage traps.  Only one ion is trapped in each trap, which is a harmonic well.  The traps are separated by a distance $L$, which is large enough to prevent disruption from the others.  The position displacement of the $i$th ion, $x_i$, is accounted with respect to the trap centre.}
\end{figure}

The configuration of our system is schematically shown in Fig.~\ref{fig:config}.  Ions are trapped in an array of harmonic storage traps, and only one ion is present in each trap. The distance between the equilibrium position of two neighbouring traps is $L$, which is sufficiently large that Coulomb coupling between the ions can be neglected.  Hence the total Hamiltonian of the system is given by
\begin{equation}\label{storage H}
\hat{H}_0= \sum_n \frac{\hat{p}_n^2}{2m} + \frac{1}{2} m \omega_0^2 \hat{x}_n^2~,
\end{equation}
where the subscript $n$ denotes the quantities belonging to the ion in the $n$th trap; $\hat{x}_n$ is the operator of the $n$th ion's location measured from the equilibrium position of the $n$th trap.  The annihilation and the creation operators of the phonon mode of the $n$th ion are given by
\begin{eqnarray}
\hat{a}_n&=&\sqrt{\frac{m\omega_0}{2 \hbar}}\hat{x}_n+i \sqrt{\frac{1}{2 \hbar m \omega_0}} \hat{p}_n\nonumber \\
\hat{a}_n^\dag&=&\sqrt{\frac{m\omega_0}{2 \hbar}}\hat{x}_n-i \sqrt{\frac{1}{2 \hbar m \omega_0}} \hat{p}_n~.
\end{eqnarray}

The ions are cooled to both the motional and electronic ground state before an input state is created.  The trapping potentials will be varied, but the potentials should return to the original form in Eq.~(\ref{storage H}) after each operation.  Thus, an operation is characterized by the transformation of the motional state in the interaction picture, i.e. 
\begin{equation}\label{S matrix state}
|\psi_I(T)\rangle =  \hat{\mathcal{S}}|\psi_I(0)\rangle~,
\end{equation}
where $\hat{\mathcal{S}}\equiv \exp(i \hat{H}_0t/\hbar) \hat{U}_S$ is the S-matrix; $\hat{U}_S$ is the evolution operator in the Schr\"{o}dinger picture.  Then the annihilation operator of a phonon mode in the interaction picture is transformed as
\begin{equation}\label{S matrix}
\hat{a}_n \rightarrow \hat{\mathcal{S}}^\dag \hat{a}_n \hat{\mathcal{S}} = \hat{U}^\dag_S \hat{a}_n \hat{U}_S e^{i \omega_0 T}~.
\end{equation}
We will omit the $n$ in future discussions of single mode operation.


\subsection{Lewis-Riesenfeld theory \label{LR theory}}

The Schr\"{o}dinger equation with a time dependent harmonic potential will frequently be encountered in future sections, i.e. 
\begin{equation}\label{TDSE}
i\partial_t |\psi\rangle = \hat{H}(t)|\psi\rangle \equiv \left( \frac{\hat{p}^2}{2m} + \frac{1}{2}m \omega^2(t) \hat{x}^2 \right) |\psi\rangle~.
\end{equation}
Lewis and Riesenfeld analyzed the problem of time varying harmonic oscillators by considering the invariant operator given by the following expression \cite{LewisJr:1969p11779, Chen:2010p11932},
\begin{equation}
\hat{\mathcal{I}}(t) = \frac{(b \hat{p} -m \dot{b} \hat{x})^2}{2m}+\frac{1}{2}m\omega_0^2 \hat{x}^2 = \hbar \omega_0\left( \hat{\mathbb{A}}^\dag(t) \hat{\mathbb{A}}(t)+\frac{1}{2}\right)~,
\end{equation} 
where a dot denotes a time derivative.  The dimensionless function $b(t)$ satisfies the auxiliary equation
\begin{equation}\label{b equation}
\ddot{b}+\omega^2(t)b-\frac{\omega_0^2}{b^3}=0~,
\end{equation} 
and the operators $\hat{\mathbb{A}}(t)$ and $\hat{\mathbb{A}}^\dag(t)$ are the raising and lowering operators of the eigenstates, $|\lambda_n, t\rangle$, of $\hat{\mathcal{I}}(t)$, i.e.
\begin{eqnarray}
\hat{\mathcal{I}}(t) |\lambda_n,t\rangle &=& \lambda_n |\lambda_n,t\rangle~, \\
\hat{\mathbb{A}}(t) |\lambda_n,t\rangle &=& \sqrt{n} |\lambda_{n-1},t\rangle~,\\ 
\hat{\mathbb{A}}^\dag(t) |\lambda_{n-1},t\rangle &=& \sqrt{n} |\lambda_n,t\rangle~,
\end{eqnarray}
with $\lambda_n$ being the corresponding eigenvalues.  The invariant operator is so called because $[\hat{\mathcal{I}}(t),\hat{H}(t)]=0$.  Hence the values of $\lambda_n$ remain unchanged, and $|\lambda_n,t\rangle$ are the solutions of Eq.~(\ref{TDSE}) during the evolution, i.e. 
\begin{equation}
\langle \lambda_m, t| \hat{H}(t) | \lambda_n, t\rangle=0~,~\textrm{if}~n\neq m~.
\end{equation}
The evolution operator of the system from time $t$ to $t'$ is given by the expression
\begin{equation}
\hat{U}(t,t')=\sum_{n=0}^\infty e^{i(n+\frac{1}{2})\left(\Theta(t)-\Theta(t')\right)} |\lambda_n,t\rangle \langle \lambda_n,t'|~,
\end{equation}
where the phase is given by
\begin{equation}\label{Theta}
\Theta(t) = - \int^t_0 \frac{\omega_0}{b^2(t'')}dt''~.
\end{equation}

When the harmonic well is static, i.e. $\omega$ is a constant, the general real solution of Eq.~(\ref{b equation}) is 
\begin{equation} \label{b solution}
b= \sqrt{\frac{\omega_0}{\omega}}\sqrt{\cosh\delta+\sinh\delta\sin(2 \omega t+\varphi)}~,
\end{equation}
where $\delta$ and $\varphi$ are constant parameters.  In our case, we are only interested in the operations where the trapping potential are steady before and after the operation, i.e.
\begin{equation}
\omega(t_i)=\omega(t_f)=\omega_0~,
\end{equation} 
where $t_i$ and $t_f$ are the starting and ending time of the operation.  For simplicity, we assume both $\delta=0$ and $\varphi=0$ at the beginning, such that $b(t_i)=1$.  In general, the values of $\delta$ and $\varphi$ have to be determined by integrating Eq.~(\ref{b equation}). The invariant operator $\hat{\mathcal{I}}(t)$ is identical to $\hat{H}(t)$ at $t=t_i$, so we have
\begin{equation}
\hat{\mathbb{A}}(t_i)=\hat{a}~.
\end{equation}
After the operation, the lowering operator becomes \cite{LewisJr:1969p11779}
\begin{equation}
\hat{\mathbb{A}}(t_f)=\eta(t_f)\hat{a}+\zeta(t_f)\hat{a}^\dag~,
\end{equation}
where
\begin{eqnarray}\label{eta}
\eta(t)=\frac{1}{2}\left(\frac{1}{b}+b-i\frac{\dot{b}}{\omega_0} \right)~,\\
\zeta(t)=\frac{1}{2}\left(\frac{1}{b}-b-i\frac{\dot{b}}{\omega_0} \right)~.\label{zeta}
\end{eqnarray}
The absolute magnitude of $\eta(t)$ and $\zeta(t)$ are 
\begin{equation}\label{eta zeta abs}
|\eta(t)|=\cosh \frac{\delta}{2}~;~|\zeta(t)|=\sinh \frac{\delta}{2}~,
\end{equation}
which the normalization condition $|\eta|^2-|\zeta|^2=1$ is satisfied automatically.

The action of the harmonic potential variation is accounted by the evolution operator, i.e. $\hat{O}=\hat{U}(t_f,t_i)$.  It is readily seen that the raising operator is related to the annihilation operator as
\begin{equation}
\hat{\mathbb{A}}(t_f)=e^{i (\Theta(t_f)-\Theta(t_i))}\hat{U}(t_f,t_i) \hat{a} \hat{U}^\dag (t_f,t_i)~.
\end{equation}
Using Eq.~(\ref{S matrix}), we deduce the transformation of annihilation operator as
\begin{eqnarray} 
\hat{a} &\rightarrow& \eta^*(t_f) e^{i (\Theta(t_f)-\Theta(t_i)+\omega_0 (t_f-t_i))} \hat{a} \nonumber \\
&&- \zeta(t_f) e^{-i (\Theta(t_f)-\Theta(t_i)-\omega_0 (t_f-t_i))} \hat{a}^\dag~. \label{LR squeeze}
\end{eqnarray}
A generalization of Lewis-Riesenfeld theory, including motion of the trap centre, is given in Appendix \ref{General HO}.

\section{single mode operations \label{single}}

\begin{figure}
\centering
\subfigure[]{~~~\includegraphics{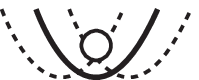}~~~~}
\subfigure[]{~~~~\includegraphics{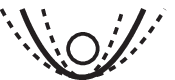}~~~~}
\subfigure[]{~~~~\includegraphics{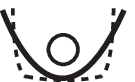}~~~}
\caption{\label{fig:operation}(a) A displacement operator is implemented by changing the trapping centre of the harmonic well.  (b) A phase-shift operator or a squeezing operator is implemented by varying the harmonic potential strength.  (c) An extra quartic potential is applied to implement the nonlinear phase gate.}
\end{figure}

Any single mode linear operator can be achieved by alternatively applying the displacement operators, phase-shift operators, and squeezing operators \cite{Lloyd:1999p11945, RMPGaussian}.  For ion trap bosonic simulations, each of these operators could be implemented by constructing specific Hamiltonians using laser interaction.  However, the accuracy and speed are limited by the validity of the Lamb-Dicke approximation (LDA), unless complicated higher order corrections are considered \cite{Wineland:1998p11782, Leibfried:2002p11781, LauJames:boson}.  We consider an alternative approach that the operators are implemented by varying the harmonic trapping potentials.  In addition, applying perturbatively a quartic potential to the storage trap can achieve a nonlinear phase gate, which is a nonlinear operator that comprise the universal set of operators.  Both harmonic potential and quartic potential can be implemented in experiments \cite{Home:2006p11926}.  All of the operations are assumed to operate from $t=0$ to $t=T$.  Summary of the operations is shown in Fig. \ref{fig:operation}.

\subsection{Displacement Operator \label{displacement operator}}

A displacement operator, $\hat{\mathcal{D}}(\alpha) = \exp (\alpha \hat{a}^\dag - \alpha^* \hat{a})$, transforms the annihilation operator as
\begin{equation}
\hat{a} \rightarrow \hat{\mathcal{D}}^\dag(\alpha) \hat{a} \hat{\mathcal{D}}(\alpha) = \hat{a}+\alpha~,
\end{equation}
where $\sqrt{2\hbar/m\omega_0}\textrm{Re}(\alpha)$ is the shift of the ion's position; $\sqrt{2\hbar m\omega_0}\textrm{Im}(\alpha)$ is the shift of the ion's momentum.  The operator can be achieved by applying two radiation fields with close frequencies, to induce a Raman transition.  If the frequency difference of the Raman fields is resonant to the red sideband of the ground electronic state, the effective Hamiltonian is then proportional to $\hat{a}+\hat{a}^\dag$ at the first order of the LDA and the rotating wave approximation (RWA), while the internal state remains at the ground state  \cite{Meekhof:1996p11999, Monroe:1996p12537, LauJames:boson}.

Another way of performing the displacement operator is to move the harmonic trap, i.e. replacing the storage trap Hamiltonian by the displacement operator Hamiltonian
\begin{equation}
\hat{H}_D=\frac{\hat{p}^2}{2m}+\frac{1}{2}m\omega_0^2 (\hat{x}-s(t))^2~;
\end{equation}
where $s(t)$ specifies the path of the trap centre .  Assume the trap centre initially located at the origin, i.e. $s(0)=0$, we require the trapping centre returns to the origin after the operation, i.e. $s(T)=0$.  After the operation, every coherent state, $|\chi\rangle$, will become
\begin{equation}
|\chi\rangle \rightarrow \Big|\left(\chi-\sqrt{\frac{m \omega_0}{2 \hbar}} \int^T_0 \dot{s}(t) \exp(i \omega_0 t) dt \right) e^{-i \omega_0 T} \Big\rangle
\end{equation}
up to a global phase that will not affect the simulation result \cite{Lau:2011p11390}.  Since the result applies to every coherent state, the annihilation operator transforms according to Eq.~(\ref{S matrix}) is
\begin{equation}\label{alpha functional}
\hat{a} \rightarrow \hat{a} - \sqrt{\frac{m \omega_0}{2 \hbar}} \int^T_0 \dot{s}(t) \exp(i \omega_0 t) dt \equiv \hat{a}+\alpha[s(t)]~,
\end{equation}
where the displacement $\alpha$ is a functional of $s(t)$.  We note that many different $s(t)$ can produce the same $\alpha$.  A method to obtain an $s(t)$ for a specific $\alpha$ is described in Appendix \ref{displacement construction}.


\subsection{Phase-Shift Operator\label{phase-shift operator}}

A phase-shift operator $\hat{\mathcal{P}}(\phi)=\exp(-i \phi \hat{a}^\dag \hat{a})$ transforms the annihilation operators as
\begin{equation}
\hat{a} \rightarrow \hat{\mathcal{P}}^\dag(\phi) \hat{a} \hat{\mathcal{P}}(\phi) = \hat{a} e^{-i\phi}.
\end{equation}
The operator can be implemented by applying a laser field to the ion, where the laser frequency is detuned from any electronic transition.  The ac Stark effect produced by the field will disturb the electronic ground state energy, and hence induces a phase shift.

Wineland \textit{et al.} \cite{Leibfried:2002p11781} implemented the phase-shift operator by applying a perturbative quadratic potential, $\epsilon \hat{x}^2$.  Under the RWA, the term $\hbar \epsilon (\hat{a} \hat{a}^\dag+ \hat{a}^\dag \hat{a})/2 m \omega_0$ dominates the perturbation and produces a phase shift.  In this approach, the validity of the RWA requires that the strength of the perturbation, $\hbar \epsilon /2 m \omega_0$, be much smaller than the  energy of a phonon $\hbar \omega_0$.  The requirement implies that the duration of a phase-shift operator, which scale as $\sqrt{m/\epsilon}$, should be much longer than $1/\omega_0$.

The operation time can be reduced to the same order as $1/\omega_0$ if the strength of the harmonic trap changes non-perturbatively, i.e. replacing the storage trap Hamiltonian by the phase-shift operator Hamiltonian
\begin{equation}
\hat{H}_P=\frac{\hat{p}^2}{2m}+\frac{1}{2}m\omega^2(t) \hat{x}^2~.
\end{equation}
The trapping frequency should be the same as that of the storage trap before and after the operation, i.e. $\omega(0)=\omega(T)=\omega_0$.  The effect of this potential can be considered analytically using the Lewis-Riesenfeld theory \cite{LewisJr:1969p11779} introduced above.  Because the amplitude of each motional Fock state remains the same after the phase-shift operator, it implies that $|\lambda_n,0\rangle = |\lambda_n, t\geq T\rangle = |n\rangle$ up to a phase, where $|\lambda_n,t\rangle$ is an eigenstate of the invariant operator at time $t$.  This criterion is equivalent to require the corresponding auxiliary function of $\omega(t)$, $b_\phi(t)$, which follows Eq.~(\ref{b equation}), to satisfy the boundary conditions
\begin{equation}
b_\phi(t \leq 0) =1~;~b_\phi(t \geq T) =1~.
\end{equation}
Then the final lowering operator will consist of only the annihilation operator, i.e. $\eta(t\geq T)=1$, $\zeta(t\geq T)=0$.  According to Eq.~(\ref{LR squeeze}), the overall effect of the harmonic potential variation will transform the annihilation operator as
\begin{equation}
\hat{a}\rightarrow \hat{a} e^{i(\Theta(T)+\omega_0 T)}~,
\end{equation}
which is obviously a phase-shift operator.

There is no unique form of $\omega(t)$ and $b_\phi(t)$ that satisfies all the above conditions, therefore we have the freedom to choose $\omega(t)$ in a manner that is convenient in practice.  Alternatively, we can initially guess a $b(t)$ and obtain the corresponding $\omega(t)$ for the experiment.  A possible choice is
$b_\phi(t) = 1-k \exp(-(t-T/2)^2/\sigma^2)~,$
where $1/\sigma \ll T$ is the characteristics time scale of the operation; $k$ is chosen to produce the desired phase shift.  There is no fundamental limitation on the magnitude of $\sigma$, so our phase-shift operator implementation can be processed indefinitely fast, which can even be faster than $1/\omega_0$ if the apparatus permits.

\subsection{Squeezing Operator  \label{squeezing operator}}

A squeezing operator $\hat{S}(g)=\exp\left((g^* \hat{a}^2 -g \hat{a}^{\dag2})/2 \right)$ transforms the annihilation operator as
\begin{equation}
\hat{a} \rightarrow \hat{S}^\dag(g) \hat{a} \hat{S}(g) = \cosh|g| \hat{a} - \frac{g}{|g|} \sinh|g| \hat{a}^\dag~.
\end{equation}
A squeezing operator is usually implemented by applying an interaction that the Hamiltonian involves second order terms in the annihilation and the creation operators, i.e. $\hat{a}^2$ and $\hat{a}^{\dag 2}$.  Such a Hamiltonian can be realized by Raman interaction with effective frequency $2\omega_0$, which is initiated by applying two radiation fields with $2\omega_0$ difference in frequency.
However, the magnitude of the potential generated by the Raman fields must be much smaller than $2 \hbar\omega_0$ to satisfy the RWA.  The operation time of the squeezing operator will then be much longer than $1/\omega_0$.  

Our approach is to use a time varying trapping potential, i.e. replacing the storage Hamiltonian by the squeezing operator Hamiltonian 
\begin{equation}
\hat{H}_S=\frac{\hat{p}^2}{2m}+\frac{1}{2}m\omega^2(t) \hat{x}^2~.
\end{equation}
The trapping frequency is required to return to that of the storage trap after the operation, i.e. $\omega(0)=\omega(T)=\omega_0$.  
The operation will transform the annihilation operator according to Eq.~(\ref{LR squeeze}), which is apparently a squeezing operator if $|\zeta(T)|\neq 0$.  The magnitude of the squeezing parameter is given by $|g|=\delta/2$ according to Eq.~(\ref{eta zeta abs}), which the $\delta$ has to be obtained by integrating the auxiliary equation Eq.~(\ref{b equation}) with the $\omega(t)$ applied in the operation.  The phase of the squeezing operator, $g/|g|$, comes from the complex nature of $\zeta(T)$ and $\eta(T)$, and the phase shift of $\left(\Theta(T)+\omega_0 T\right)$ in Eq.~(\ref{LR squeeze}).  The effect of this operation can be accounted by analytically solving the time dependent harmonic oscillator, so the operation time is not limited by the validity condition of the RWA.

The applied potentials of $\hat{H}_S$ and $\hat{H}_P$ are both time varying harmonic wells; the only difference is the time variation of the trapping frequency, $\omega(t)$.  Unless $\omega(t)$ is specially designed as shown in Sec. \ref{phase-shift operator}, the operation of varying the harmonic potential would be a squeezing operator with some squeezing parameter crucially depending on $\omega(t)$.  The $\omega(t)$ that generates a specific squeezing parameter can be obtained by the method detailed in Appendix \ref{squeezing construction}.

\subsection{Nonlinear Operator \label{nonlinear operator}}

Nonlinear operators transform an annihilation operator to an operator involving quadratic and higher order terms in $\hat{a}$ and $\hat{a}^\dag$.  It can be achieved by applying a Hamiltonian that is at least third order of $\hat{a}$ and $\hat{a}^\dag$.  One implementation is to exert a radiation field that is resonant to high order sideband frequencies.  For example, a Raman field with the effective frequency $3 \omega_0$ can provide a Hamiltonian scaling as $(\hat{a}^3-\hat{a}^{\dag 3})$ at the third order expansion of the Lamb-Dicke parameters \cite{Meekhof:1996p11999}.  The major problem of this approach is that the validity of both the LDA and the RWA have to be satisfied, so the undesired Hamiltonian are suppressed.  The consequence is that the power of the radiation field is constrained, which limits the speed of the operation.  Nonetheless, our architecture can facilitate this laser-mediated nonlinear operator, because the mode spectrum is simplified since only one bosonic mode is exhibited in each trap.

A nonlinear operator can also be implemented by switching on an additional quartic potential, i.e. 
\begin{equation}
\hat{H}_4(t) = \hat{H}_0+ \hat{V}_4(t) \equiv \hat{H}_0+ \mathcal{F}(t) \hat{x}^4~.
\end{equation}
In the interaction picture with respect to $\hat{H}_0$, the quartic potential becomes
\begin{eqnarray}
\hat{V}_4^{I}&=&\frac{\mathcal{F}(t) \hbar^2}{(2 m \omega_0)^2} \Big[(6 \hat{a}^{\dag 2} \hat{a}^2 +12 \hat{a}^\dag \hat{a} +3) \nonumber \\
&& + a^{\dag 2}(4 \hat{a}^\dag \hat{a}+6)e^{i 2\omega_0 t} + \hat{a}^{\dag 4} e^{i 4\omega_0 t} + \textrm{h.c.} \Big] ~.
\end{eqnarray}
If the variation of $\mathcal{F}(t)$ is slow enough, the off-resonant terms can be eliminated by the RWA; 
the only effective terms are
\begin{equation}
\hat{H}_N \equiv \frac{\mathcal{F}(t) \hbar^2}{(2 m \omega_0)^2} (6 \hat{a}^{\dag 2} \hat{a}^2 +12 \hat{a}^\dag \hat{a} +3)~.
\end{equation}
Applying the quartic potential from $t=0$ to $T$, the S-matrix of the operation in the Schr\"{o}dinger picture is given by
\begin{equation}\label{S nonlinear}
\hat{\mathcal{S}}_4 = e^{-i \mu(T) (6 \hat{a}^{\dag 2} \hat{a}^2 + 12 \hat{a}^{\dag} \hat{a})}~,
\end{equation}
where
\begin{equation}
\mu(t)= \int^T_0 \frac{\mathcal{F}(t') \hbar}{(2 m \omega_0)^2} dt'~.
\end{equation}
We have neglected the unimportant global phase, and have employed the fact that $[\hat{H}_0,\hat{H}_N]$.

The speed of the operation is mainly determined by the validity of the RWA.  According to Ref. \cite{Gamel:2010p12539}, applying the RWA is to collect the leading order terms in a series expansion of time-averaged Hamiltonians.  The sufficient condition for a valid series expansion is that the largest eigenvalue of $\hat{H}_N/\hbar$ should be much smaller than the off-resonant frequencies, which are multiples of $\omega_0$ in our case.  Although $\hat{H}_N/\hbar$ has unbounded eigenvalues, the series expansion is still valid if the maximum phonon number, $n_\textrm{max}$, in each mode is small.  To estimate the RWA validity condition,
we approximate $\mathcal{F}(t) \hbar^2/(2 m \omega_0)^2$ by $\hbar\mu(T)/T$ because $\hat{V}_4$ is slowly varying.  The maximum eigenvalue of $\hat{H}_N$ in our simulation is hence $n_\textrm{max}^2 \mu(T)/T$, which gives a valid RWA when
\begin{equation} \label{RWA condition}
\frac{n_\textrm{max}^2 \mu(T)}{\omega_0 T} \ll 1~.
\end{equation}


Arbitrary evolution of a single phonon mode can be realized by repeatedly applying the linear operators and the nonlinear phase-shift operator \cite{Lloyd:1999p11945}.  The idea is to consider the intertwinement of the infinitesimal evolution of two Hamiltonian $\hat{\mathcal{Y}}$ and $\hat{\mathcal{Z}}$, i.e.
\begin{equation}\label{concatenate}
e^{i\hat{\mathcal{Y}}\delta t/\hbar}e^{i \hat{\mathcal{Z}}\delta t/\hbar}e^{-i \hat{\mathcal{Y}}\delta t/\hbar}e^{-i\hat{\mathcal{Z}}\delta t/\hbar} = e^{i[\hat{\mathcal{Y}},\hat{\mathcal{Z}}]\delta t^2/\hbar^2} + O(\delta t^3)~.
\end{equation}
If the $O(\delta t^3)$ terms are neglected, the resultant operation of this sequence is effectively the evolution of a new Hamiltonian $[\hat{\mathcal{Y}},\hat{\mathcal{Z}}]$.  Evolution of new classes of Hamiltonian will be generated by using this trick again; the evolution of the desired Hamiltonian is eventually produced.  

The corresponding Hamiltonian of a linear operator consists of second or lower orders of $\hat{a}$ and $\hat{a}^\dag$.  For our selection of linear operators, the Hamiltonian of $\hat{\mathcal{D}}(\alpha)$, $\hat{\mathcal{P}}(\phi)$, $\hat{S}(g)$ are respectively proportional to $\hat{a} $, $\hat{a}^\dag \hat{a}$, $\hat{a}^2$ and their Hermitian conjugates.  Hamiltonians with higher than second order of $\hat{a}$ and $\hat{a}^\dag$ cannot be generated by intertwining the linear operators, therefore the nonlinear phase-shift operator is required to implement UBS.


\section{Two-mode Operation \label{phonon beam splitter}}

In Sec. \ref{single}, we examined the techniques to realize single mode operations.  We now turn our attention to mixing two modes, analogous to a beam-splitter in optics.
A linear beam splitting operation, $\hat{\mathcal{B}}(\theta, \phi)$, transforms two boson modes as
\begin{eqnarray}
\hat{a}_1 \rightarrow \hat{\mathcal{B}}^\dag \hat{a}_1 \hat{\mathcal{B}} &=& \cos\frac{\theta}{2} \hat{a}_1 +  i\sin \frac{\theta}{2} e^{i\phi} \hat{a}_2~; \nonumber \\
\hat{a}_2 \rightarrow \hat{\mathcal{B}}^\dag \hat{a}_2 \hat{\mathcal{B}} &=& i\sin \frac{\theta}{2} e^{-i\phi} \hat{a}_1 +  \cos \frac{\theta}{2} \hat{a}_2~. \label{beam splitter}
\end{eqnarray}
A 50:50 beam splitter corresponds to the case $\theta=\pi/2$.  

\begin{figure}
\centering
\includegraphics{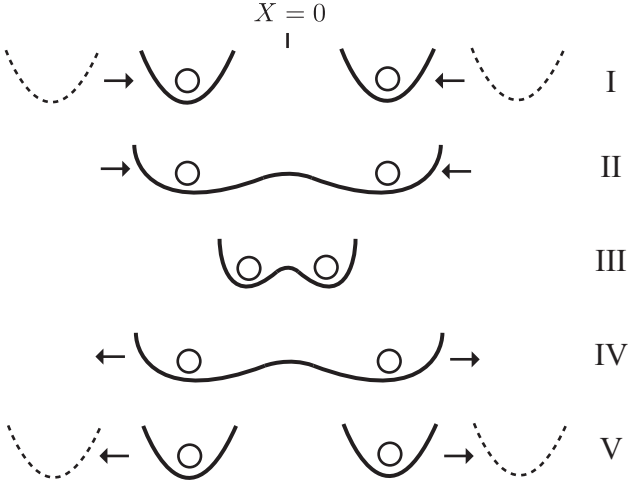}
\caption{\label{fig:bs} Displacement of ions and variations of potentials during a beam splitter operation.  The origin is defined as the mid-point between two storage traps.  Step I, ions are transported by harmonic well from storage traps to pick-up position.  Step II, double well is switched on to pick-up ions.  Step III, separation of double well shrinks.  Step IV, the separation of double well extends and brings the ions back to pick-up position.  Step V, double well is switched off, and the harmonic wells pick-up ions and bring them to storage traps.}
\end{figure}

The whole beam splitter process is summarized schematically in Fig.~\ref{fig:bs}.  Displacing harmonic wells are applied from $t=-T/2$ to transport the ions from the storage traps to the pick-up distance and then switched off at $t=-T'/2$, a double well potential is immediately switched on to relay the transportation.  The separation of the double well shrinks and then expands; the two ions are brought to proximity and then separated.  The two encoded phonon modes interact via the Coulomb interaction between the ions.  
The double well potential finally separates the ions to the pick-up distance and then switched off at $t=T'/2$, two moving harmonic wells are switched on immediately to transport the ions back to the storage traps at $t=T/2$.

In this section we set the origin of position, $X=0$, is the mid-point between the two storage traps.  We assume the system is both spatially and dynamically symmetric about the origin.  To simplify the discussion, we will separate the evolution of the classical motion and the quantum fluctuations by defining 
\begin{equation}\label{var separate}
\hat{X}_i \equiv \bar{X}_i + \hat{q}_i~;~\hat{P}_i \equiv \bar{P}_i + \hat{\pi}_i
\end{equation}
where the subscripts $i=1,2$ denote the ions involved in the beam splitter operation; $\hat{X}$ and $\hat{P}$ are the total position and momentum operator; $\bar{X}_i$ and $\bar{P}_i=m\dot{\bar{X}}_i$ are the classical position and momentum of the $i$th ion;  $\hat{q}_i$ and $\hat{\pi}_i$ are the operators accounting for the quantum fluctuation of position and momentum about $\bar{X}_i$ and $\bar{P}_i$ respectively.  The aim of our beam splitter is to transform the quantum fluctuations of the two ions according to Eq.~(\ref{beam splitter}), whereas the ions will be classically stationary at the storage traps before and after the operation, i.e. 
\begin{eqnarray}
\bar{X}_1(-\frac{T}{2})=\bar{X}_1(\frac{T}{2})=-\frac{L}{2}~;~\bar{X}_2(-\frac{T}{2})=\bar{X}_2(\frac{T}{2})=\frac{L}{2}~;~~\\
\bar{P}_1(-\frac{T}{2})=\bar{P}_1(\frac{T}{2})=\bar{P}_2(-\frac{T}{2})=\bar{P}_2(\frac{T}{2})=0~.~~
\end{eqnarray}
At the storage traps, the quantum fluctuations are the same as the phonon quadratures defined in Eq.~(\ref{storage H}), i.e.
\begin{equation}
\hat{q}_i \big|_{\textrm{storage}} = \hat{x}_i~;~\hat{\pi}_i \big|_{\textrm{storage}} = \hat{p}_i~.
\end{equation}

The core component of our phonon beam splitter is a double well potential with varying well separation.  
It can be constructed by a quartic potential, $A(t) X^4$, and a harmonic potential, $B(t) X^2$, which can be implemented in experiments \cite{Home:2006p11926}.  The evolution of the motional state follows the Schr\"{o}dinger equation $i \hbar \partial_t |\psi\rangle = \hat{H}_{BS}(t) |\psi\rangle$, where
\begin{eqnarray}\label{BS H1}
\hat{H}_{BS}(t)&=&\frac{\hat{P}_1^2}{2m} + \frac{\hat{P}_2^2}{2m} + B(t) (\hat{X}_1^2 + \hat{X}_2^2) \nonumber \\&&
+ A(t)(\hat{X}_1^4 + \hat{X}_2^4) +\frac{e^2}{4 \pi \epsilon_0 (\hat{X}_2-\hat{X}_1)}~;
\end{eqnarray}
$|\psi\rangle$ is the total wave function.  We have assumed the ions cannot tunnel pass each other due to the strong Coulomb repulsion, so $X_2>X_1$ is always true.  

In terms of the variables in Eq.~(\ref{var separate}), the Schr\"{o}dinger equation becomes
\begin{equation}
i \hbar \partial_t |\tilde{\psi}\rangle = (\mathcal{H}_0 + \hat{\mathcal{H}}_1 + \hat{\mathcal{H}}_B)  |\tilde{\psi}\rangle~,
\end{equation}
where $\mathcal{H}_0$ and $\hat{\mathcal{H}}_1$ collect the terms with the zero and first order quantum fluctuations, and $\tilde{H}_B$ contains the rest; $|\tilde{\psi}\rangle$ is the state of the quantum fluctuations.  The first term $\mathcal{H}_0 = \bar{P}_1^2/2m + \bar{P}_2^2/2m + A(t) (\bar{X}_1^4+\bar{X}_1^4)+B(t) (\bar{X}_1^2+\bar{X}_1^2) + e^2/4 \pi \epsilon_0 (\bar{X}_2-\bar{X}_1)$ is the total mechanical energy of the system; it only contributes to an unimportant global phase.  The second term $\hat{\mathcal{H}}_1$ vanishes if the classical equations of motion are satisfied, i.e.
\begin{eqnarray}
\dot{\bar{X}}_i&=& \bar{P}_i/m~, \nonumber \\
\dot{\bar{P}}_i&=&-4A(t)\bar{X}_i^3-2B(t)\bar{X}_i+\frac{e^2 \bar{X}_i}{4 \pi \epsilon_0 |\bar{X}_i|(\bar{X}_2-\bar{X}_1)^2}~.\nonumber
\end{eqnarray}
Because of the symmetry, we have $\bar{X}_1(t)=-\bar{X}_2(t)$ and $\bar{P}_1(t)=-\bar{P}_2(t)$.  The classical separation between the ions is defined as $r=\bar{X}_2-\bar{X}_1$, then the equation of motion reduces to the following:
\begin{equation}\label{EOM}
\ddot{r}=-\frac{A(t)}{m}r^3-\frac{2B(t)}{m}r+\frac{e^2}{4 \pi \epsilon_0 m r^2}~.
\end{equation}

If the quantum fluctuation of position is much smaller than the separation of ions, i.e. $\sqrt{\langle \delta \hat{q}^2 \rangle}/r \ll 1$, then $\hat{\mathcal{H}}_B$ can be approximated by a quadratic Hamiltonian, viz
\begin{eqnarray}\label{BS H2}
\hat{\mathcal{H}}_B \approx \hat{\mathcal{H}}_2 &=& \frac{\hat{\pi}_1^2}{2m} + \frac{\hat{\pi}_2^2}{2m} + \left(\frac{3}{2}A(t) r^2+ B(t)\right) (\hat{q}_1^2 + \hat{q}_2^2) \nonumber \\&&
+\frac{e^2 (\hat{q}_2-\hat{q}_1)^2}{8 \pi \epsilon_0 r^3}~.
\end{eqnarray}

Instead of staying in the bases of individual ions, it is advantageous to consider the centre-of-mass mode (+ mode) and breathing mode (- mode).  
The position and momentum operators are defined as
\begin{equation}
\hat{q}_\pm =\frac{\hat{q}_2\pm\hat{q}_1}{\sqrt{2}}~;~\hat{\pi}_\pm =\frac{\hat{\pi}_2\pm\hat{\pi}_1}{\sqrt{2}}~.
\end{equation}
In the new basis, Eq.~(\ref{BS H2}) can be decoupled as two harmonic oscillators,
\begin{equation}
\hat{\mathcal{H}}_2=\frac{\hat{\pi}_+^2}{2m}+\frac{1}{2}m \omega_+^2(t) \hat{q}_+^2+\frac{\hat{\pi}_-^2}{2m}+\frac{1}{2}m \omega_-^2(t) \hat{q}_-^2~,
\end{equation}
where the mode frequencies are
\begin{eqnarray}\label{omega+}
\omega_+(t) = \sqrt{\frac{3 A(t)}{m} r^2+ \frac{2B(t)}{m}}~; \\
\label{omega-}
\omega_-(t) = \sqrt{\omega_+^2(t)+\frac{e^2}{2 \pi \epsilon_0 m r^3}}~.
\end{eqnarray}
The annihilation operators of the modes are defined as
\begin{equation}
\hat{\mathcal{A}}_\pm=\sqrt{\frac{m \omega_\pm}{2 \hbar}} \hat{q}_\pm + i \sqrt{\frac{1}{2 \hbar m \omega_\pm}} \hat{\pi}_\pm~.
\end{equation}


Changing the magnitude of the double well will result in either squeezing or phase-shifting the + mode and the - mode.  Because there is no excitation after a beam splitting, the double well operation should give only a phase shift.  For simplicity, we assume the quartic and the harmonic potentials are adjusted to produce a constant $\omega_+$, i.e.
\begin{equation}\label{constant omega+}
\frac{3 A(t)}{m} r^2(t)+ \frac{2B(t)}{m} = \omega^2_+(t) = \omega^2_0~.
\end{equation}
According to Eq.~(\ref{LR squeeze}), the + mode remains unchanged after the operation, i.e. $\hat{\mathcal{A}}_+ \rightarrow \hat{\mathcal{A}}_+$.  

We require the double well operation is merely a phase-shift operator on the breathing mode, i.e. the annihilation operator transforms as $\hat{\mathcal{A}}_- \rightarrow e^{-i\theta}\hat{\mathcal{A}}_-$, where $\theta =  (-\Theta(T)-\omega_0 T)$ according to Eq.~(\ref{LR squeeze}).  
The phonon modes of individual ions then transform as 
\begin{eqnarray}
\hat{\mathcal{A}}_1 \rightarrow \frac{1}{\sqrt{2}}(\hat{\mathcal{A}}_+ - e^{-i\theta} \hat{\mathcal{A}}_-) = \cos \frac{\theta}{2} \hat{\mathcal{A}}_1 + i \sin\frac{\theta}{2} \hat{\mathcal{A}}_2~; \\
\hat{\mathcal{A}}_2 \rightarrow \frac{1}{\sqrt{2}}(\hat{\mathcal{A}}_+ + e^{-i\theta} \hat{\mathcal{A}}_-) = i \sin \frac{\theta}{2} \hat{\mathcal{A}}_1 + \cos \frac{\theta}{2} \hat{\mathcal{A}}_2~,
\end{eqnarray}
which is a beam splitter operation.  We have neglected the unimportant global phase $e^{-i \theta/2}$, and the phase $\phi$ in Eq.~(\ref{beam splitter}) can be rectified by applying local phase-shift operators.  

The pick-up distance should be large enough that $\omega_-$ and $\omega_+$ should be roughly the same according to Eq.~(\ref{omega-}), i.e. $\omega_-(T'/2)=\omega_-(T'/2)=\omega_0$.
To produce only a phase shift on - mode, the $\omega_-(t)$ should produce an auxiliary function $b(t)$ that satisfies $b(-T'/2)=b(T'/2)=1$.  We can control the strength of the quartic and harmonic potentials, $A(t)$ and $B(t)$, to produce an appropriate $\omega_-(t)$ while preserving $\omega_+(t)$.  The appropriate time variations of $A(t)$ and $B(t)$ exist and are not unique; they can be chosen in a manner that is convenient in practice.   In Appendix \ref{bs construction}, we suggest a method to obtain $A(t)$ and $B(t)$ from a speculative $b(t)$.


Before and after the double well operation, the ions are transported back and forth between the storage traps and the pick-up distance (step I and V).  If both the transporting harmonic potentials and the double well potential can be switched on and off quickly, the pick-up process can be conducted smoothly that the phonon states will not be disturbed.  The pick-up distance is arbitrary; it can be chosen in a manner that is favourable in experiments.


The ions' classical velocity in the double well operation is determined by the choice of $A(t)$ and $B(t)$.
The velocity at the pick-up distance, $\dot{\bar{X}}_i (-T'/2)$ and $\dot{\bar{X}}_i (T'/2)$, is obtained by integrating the equation of motion Eq.~(\ref{EOM}) and the condition of a vanishing velocity at the turning point, i.e. $\dot{r}(0)=0$.   In step I, the transporting harmonic wells should increase the classical velocity of the ions from 0 at the storage trap to $\dot{\bar{X}}_i (-T'/2)$ at the pick-up distance.  Similarly in step V, the transporting harmonic wells should reduce the classical velocity from $\dot{\bar{X}}_i (T'/2)$ at the pick-up distance to 0 at the storage traps.
The classical position and momentum of the ions in the transportation process depend on the trap centre, $\xi_i(t)$, of the transporting harmonic well, which the Hamiltonian is
$\hat{H}_T = \hat{P}^2_i/2m + m \omega_0^2 (\hat{X}_i - \xi_i(t) )^2/2$.  Using the variables in Eq.~(\ref{var separate}), the classical equation of motion is 
\begin{equation}
m\ddot{\bar{X}}_i(t)=\omega_0^2 \left(\xi_i(t)-\bar{X}_i(t) \right)~,
\end{equation}
where the exact solution can be found in Ref. \cite{Lau:2011p11390}.  Appropriate $\xi_i(t)$, which produces the $\bar{X}_i$ and $\bar{P}_i$ that match the boundary conditions at $t=-T/2, -T'/2, T'/2, T/2$, can be obtained by the inverse-engineering method presented in Ref. \cite{Torrontegui:2011p9810}.  The evolution of the quantum fluctuations in the transportation follows the Hamiltonian
\begin{equation}
\hat{\mathcal{H}}_T= \frac{\hat{\pi}_i^2}{2m}+\frac{1}{2}m\omega_0^2 q_i^2~.
\end{equation}
Obviously, the phonon states will not be disturbed by $\hat{\mathcal{H}}_T$ and hence the transportation.  All in all, the operation from step I to step V realizes a phonon beam splitter, i.e. Eq.~(\ref{beam splitter}), on the phonon modes in neighbouring storage traps.



Now let us consider the errors in the beam splitter operation.  The transportation in step I and V would cause small error if the harmonic well is sufficiently accurate.  The pick-up process is assumed to be fast enough that does not cause significant error.  
Thus, most of the error is expected to come from the double well process.  The main problem is that the anharmonic terms in $\hat{\mathcal{H}}_B$, which are the terms with at least third order of quantum fluctuations and are not covered by $\hat{\mathcal{H}}_2$, would produce unwanted excitation or phase shift.  The error is expected to be magnified if more phonons are involved in the beam splitter, because the significance of the quantum fluctuation is characterized by $\sqrt{\langle \hat{q}^2 \rangle}/r$, which increases as phonon number.  A faster operation also causes more significant errors, because some of the anharmonic terms would be suppressed by RWA if the operation is slow.  The evolution of phonon states is calculated in Appendix \ref{bs accuracy} when some anharmonic terms are included in the Hamiltonian.  We find that a beam splitter can be conducted as fast as $3~\mu$s if there are less than 8 phonons involved in each $^{40}$Ca$^+$ ion trap.  

The technique of double well operation can also be applied to split adiabatically a pair of ions in a single trap, details are provided in Appendix \ref{ion separation}.

\section{Initialization and readout \label{initialization}}

Manipulation of trapping potentials is sufficient for initializing arbitrary Gaussian phonon states, which can be created by applying linear operators to the ground motional states.  However, non-Gaussian phonon states, such as Fock states and the Schr\"{o}dinger's cat state, have to be generated by laser interaction.  When comparing with the previous UBS proposal that incorporates a chain of ions in a single trap \cite{Wineland:1998p11782, LauJames:boson}, our architecture simplifies the laser-mediated state initialization. Because the ions are separately trapped, sophisticated techniques to prevent the laser operation from disturbing other ions, such as using composite pulses and shielding, are no longer necessary.

Information about the phonon states can be extracted by
the three measurement schemes suggested in Ref. \cite{LauJames:boson}: using adiabatic transfer, post-selection techniques, and multiple electronic states.  
The adiabatic transfer is achieved by exerting a Raman pulse whose frequency is slowly increased from lower to higher than a sideband frequency \cite{Cirac:1994p11995}.
The phonon-dressed electronic states transfer as $|g~n\rangle \leftrightarrow |e~(n-1)\rangle$ while $|g~0\rangle$ remains unchanged.  The fluorescence measurement scheme in ion trap quantum computing is conducted after one round of adiabatic transfer, i.e. by applying a strong laser pulse resonant to the transition frequency between $|g\rangle$ and some other unstable electronic state of the ion; significant fluorescence is detected if the electron is in the state $|g\rangle$.  A positive measurement outcome is produced if the original phonon state is $|0\rangle$.  This procedure is equivalent to the non-distinguishing number detector in optical experiments.
If the frequency detuning and the Rabi frequency of the Raman field are both tuned, an adiabatic transfer can be achieved for a single mode with 0.99 fidelity in as fast as $80~\mu$s \cite{LauJames:boson, book:AllenEberly}.  

If more rounds of adiabatic transfer and carrier pulses are applied, a fluorescence measurement can produce the projection-value measure (PVM), $\{\sum_{n=0}^m|n\rangle\langle n|, \sum_{n=m+1}^\infty|n\rangle\langle n| \}$, which can distinguish if the phonon population is more or less than an integer $m$.  If there is another meta-stable electronic state, then any Fock state $|m\rangle$ can be deterministically measured by the PVM $\{|m\rangle\langle m|, \mathbb{I}-|m\rangle\langle m| \}$, where $\mathbb{I}$ is the identity matrix.  Incorporating with the displacement operator, any coherent state $|\alpha\rangle$ can also be deterministically distinguished by the PVM $\{|\alpha\rangle\langle \alpha|, \mathbb{I}-|\alpha\rangle\langle \alpha| \}$ \cite{Poyatos:1996p11928}.  To the best of our knowledge, detectors with these PVM have not yet been developed in optical experiment.

The post-selection measurement method is applicable only if a negative fluorescence measurement causes negligible distortion on the resultant phonon state.  Post-selection quantum information protocols, such as the linear optics entanglement gate \cite{Knill:2001p8225}, can be achieved by using this measurement method.  A sideband pulse is applied to transit some unwanted superposition to $|e\rangle$, while the desired outcome state remains in $|g\rangle$.  The unwanted superpositions will be removed if the fluorescence measurement with respect to $|e\rangle$ gives a negative result.  Repeating this process will increase the amplitude of the desired outcome state in the residual phonon state.  

If the maximum number of phonons in a mode is known as $n_{\textrm{max}}$, and there are $n_{\textrm{max}}$ meta-stable electronic states available for manipulation, each Fock state with phonon number smaller than or equal to $n_{\textrm{max}}$ can be projectively measured, i.e. phonon number distinguishing measurement.  The principle is to associate each phonon Fock state to an electronic state; the fluorescence measurement is then conducted on the electronic states one by one.  The sequence of pulses employed to measure a motional state with $n_{\textrm{max}}=2$ is given in \cite{LauJames:boson}.

In all the three measurement schemes, our architecture is more favourable than the previous proposals in Ref. \cite{Wineland:1998p11782, LauJames:boson} that traps a chain of ions in a single harmonic potential.  Because the ions are individually trapped, the recoil of an ion after fluorescence measurements do not distort other phonon modes.  The spectral distribution of resonance is also simplified because only one mode is present in each trap; the speed of sideband transition is hence increased as a stronger pulse can be used without the concern of accidental mode mixing.

\section{Conclusion \label{conclusion}}

We have described a possible architecture to implement the universal bosonic simulator using individually trapped ions.  The excitation of an ion's quantized motion can simulate a bosonic mode.  Linear single mode operations can be realized by changing the strength and the trapping centre of the harmonic potential at the storage trap.  Nonlinear phase-shift operator can be implemented by perturbatively exerting a quartic potential.  Linear beam splitter is implemented by a double well potential with varying separation; the Coulomb interaction between the ions ensues from the interaction between phonons.  By applying the operators alternatively, arbitrary bosonic evolution can be effectively simulated \cite{Lloyd:1996p10335, Lloyd:1999p11945}.

All linear and nonlinear operators can be conducted without laser interaction, hence the speed of the operations are not limited by the Lamb-Dicke approximation.  However, all the laser-mediated techniques of the previous proposals in Ref. \cite{Wineland:1998p11782, LauJames:boson}, which employ a chain of ion in a single trap, are applicable to our architecture.  Because only one phonon mode is associated to each trap, the resonant frequency spectrum is much simpler than a trap with multiple ions.  The requirement of the rotating wave approximation is also less stringent due to the absence of accidental mode mixing.  In addition, measuring a phonon mode by laser keeps the other phonon modes undistorted.

Provided that the quality and the controllability of the harmonic potential are fine enough, the operation time of the single mode linear operators has no fundamental limit, and a two-mode operator can be implemented within tens of $1/\omega_0$ if the phonon number in each interacting mode is less than 10.
  Although the speed of the nonlinear operator is relatively slower, various interesting bosonic phenomena can be investigated using only linear operators.  
  
Recently, Aaronson and Arkhipov \cite{Aaronson:2010p11906} proposed that if there exists a classical algorithm that efficiently samples the probability distribution of a linear bosonic network, then the polynomial hierarchy would collapse to the third level, which is generally believed to be impossible in computer science \cite{Phierachy}.  In other words, an approximate bosonic sampler should be a machine having \textit{post-classical computing power}; building such a machine can verify the power of quantum computer.  Aaronson and Arkhipov suggest that a boson sampler involving $n=10$ to $50$ bosons and about $\sim n^2$ to $\sim n^5 \log n$ modes suffice to achieve the goal.  The scalability, speed, theoretical quality, and measurement flexibility of our ion trap bosonic simulator architecture offers the possibility of demonstrating such post-classical computation in the not too distant future.

We thank Marcos Villagra, Christian Weedbrook, and Andrew White for useful comments.  The authors would like to acknowledge support from the NSERC CREATE Training Program in Nanoscience and Nanotechnology.

\appendix

\section{Generalized solution of harmonic oscillator \label{General HO}}

We have introduced the method to individually implement the displacement operator and the squeezing operator.  In fact, both operators can be implemented in one operation, which employs a harmonic trap with both the strength and the centre of the trap are varying, i.e.
\begin{equation}\label{GHO}
i\hbar\partial_t |\psi(t)\rangle =\left(\frac{\hat{p}^2}{2m}+\frac{1}{2}m\omega^2(t) (\hat{x}-s(t) )^2\right)|\psi(t)\rangle~;
\end{equation}
where $|\psi(t)\rangle$ is the solution of the equation.  This generalized harmonic oscillator has been investigated in the context of the evolution of number states \cite{LO:1991p12938}, and the variation of quadratic operators in the Heisenberg picture \cite{Kim:1996p12936}.  We here formulate the previous results to fit in our purpose of bosonic simulation, i.e. to consider the evolution of the annihilation operator in the interaction picture according to Eq.~(\ref{S matrix}).

We define the displaced state as
\begin{equation}
|\chi(t)\rangle = \hat{\mathcal{D}}^\dag(\beta(t) e^{-i \omega_0 t}) |\psi(t)\rangle~,
\end{equation}
where $\hat{\mathcal{D}}$ is the displacement operator; the displacement is given by the expression
\begin{equation}
\beta(t) e^{-i \omega_0 t} = \sqrt{\frac{m \omega_0}{2 \hbar}} x_c(t) + i \sqrt{\frac{1}{2 \hbar m \omega_0}} p_c(t)~.
\end{equation}
If $x_c$ and $p_c$ are respectively the classical position and momentum of the ion, i.e. they obey the classical equation of motion
\begin{equation}\label{general EOM}
\dot{x}_c(t) = p_c(t)/m~;~\dot{p}_c(t)=-m\omega^2(t) (x_c(t) - s(t) )~,
\end{equation}
then the displaced state follows the equation
\begin{equation}
i \hbar \partial_t |\chi(t)\rangle = \left(\frac{\hat{p}^2}{2m}+\frac{1}{2}m\omega^2(t) \hat{x}^2\right)|\chi(t)\rangle~,
\end{equation}
which can be solved by the Lewis-Riesenfeld theory.  An unimportant global phase has been neglected in the above derivation.

The operation runs from $t=0$ to $T$.  The trap centre and strength should return to that of the storage trap at $t=T$.  The evolution operator of the displaced state in the Schr\"{o}dinger picture, $\hat{U}_{\chi,S}$, transforms the annihilation operator as
\begin{equation} 
\hat{U}^\dag_{\chi,S}\hat{a}\hat{U}_{\chi,S} = \eta^*(T) e^{i \Theta(T)} \hat{a} - \zeta(T) e^{-i \Theta(T)} \hat{a}^\dag~.
\end{equation}
The evolution operator of  $|\psi(t)\rangle$ is related to $\hat{U}_{\chi,S}$ by
\begin{equation}
\hat{U}_S(t) = \hat{\mathcal{D}}(\beta(t) e^{-i \omega_0 t}) \hat{U}_{\chi,S}(t)~.
\end{equation}
Using Eq.~(\ref{S matrix}), we can deduce the transformation of the annihilation operator in the interaction picture is
\begin{eqnarray} 
\hat{a} &\rightarrow& \eta^*(T) e^{i (\Theta(T)+\omega_0 T)} \hat{a} \nonumber \\
&&- \zeta(T) e^{-i (\Theta(T)-\omega_0 T)} \hat{a}^\dag + \beta(T)~.
\end{eqnarray}
If $b(t)$ is known, $\beta(T)$ can be obtained as \cite{Kim:1996p12936}
\begin{eqnarray}
\beta(T) &=& i \sqrt{\frac{m}{2\hbar \omega_0}} \int^T_0 b(t) \omega^2(t) s(t) e^{-i \Theta(t)}dt \nonumber \\
&&\times\left(\eta^*(T)e^{i (\Theta(T)+\omega_0 T)}+\zeta(T)e^{-i (\Theta(T)-\omega_0 T)} \right).~~~~~
\end{eqnarray}

The total operation is obviously a squeezing operator followed by a displacement operator shifting the classical position and momentum of the ion.  The squeezing parameter is exactly the same as that of the centre-fixed squeezing operator in Sec. \ref{squeezing operator} for the same $\omega^2(t)$.  A desired operation can be constructed by: first obtaining a $\omega^2(t)$ that produce the desired squeezing parameter; then finding a $s(t)$ that can produce the desired displacement $\beta(t)$ following the equation of motion Eq.~(\ref{general EOM}).

\section{Construction of displacement operator \label{displacement construction}}

The variation of the harmonic trap centre, $s(t)$, that produces a specific displacement $\alpha_0$ can be obtained systematically by the inverse engineering method in Ref. \cite{Torrontegui:2011p9810}.  We here present a simpler method that employs the linearity of displacements and paths.  First of all, two arbitrary paths are speculated, $s_1(t)$ and $s_2(t)$; both paths satisfy the boundary conditions $s(0)=s(T)=0$.  According to Eq.~(\ref{alpha functional}), the paths will produce two displacements, $\alpha_1 \equiv \alpha[s_1]$ and $\alpha_2 \equiv \alpha[s_2]$. We require $\alpha_1$ and $\alpha_2$ are not scaled by a real number, otherwise another path $s_3(t)$ has to be speculated.  If the requirement is satisfied, there must exist two parameters, $\gamma_1$ and $\gamma_2$, such that
\begin{equation}
\alpha_0=\gamma_1 \alpha_1+ \gamma_2 \alpha_2~.
\end{equation}
Then the path $s(t) = \gamma_1 s_1(t) +\gamma_2 s_2(t)$ will give the desired displacement $\alpha_0$.

\section{Construction of squeezing operator \label{squeezing construction}}

For a particular $\omega(t)$, the auxiliary function $b(t)$ should behave as Eq.~(\ref{b solution}) at $t\geq T$ after integrating Eq.~(\ref{b equation}) with the boundary condition $b(t \leq 0)=1$.  To construct a squeezing operator with the desired squeezing operator $g$, we have to find an $\omega(t)$ that generates the parameter $\delta=2 |g|$ for $b(t\geq T)$.  Such a condition can be satisfied by a wide range of $\omega(t)$; a particular $\omega(t)$ can be obtained inversely from a constructed $b(t)$.  An example is 
\begin{equation}
b_S(t) = \sqrt{\cosh \delta + \sinh \delta \sin(2 \omega t) }h(t) + (1-h(t))~,
\end{equation}
where $h(0)=0$ and $h(T)=1$.  
$\omega(t)$ should be continuous before and after the operation, so $b_S(t)$, $\dot{b}_S(t)$, and $\ddot{b}_S(t)$ have to be continuous at $t=0$ and $t=T$.  For instance, $h(t)=10(t/T)^3-15(t/T)^4+6(t/T)^6$ meets the requirement.  The time variation of $\omega(t)$ can be obtained by inputting $b_S(t)$ into Eq.~(\ref{b equation}).


So far we have neglected the phase of the squeeze operator.  In the construction described above, the phase of the operation originates in the phase of $\zeta(T)$ and $\eta(T)$, as well as the phase shift in Eq.~(\ref{LR squeeze}), i.e. $\exp\left(i (\Theta(T)+\omega_0 T)\right)$.  
The $\omega(t)$ is the unique factor determining the total phase, which is in general different from the desired value $g/|g|$.  The phase is possible to be rectified by first calculating the total phase shift generated from our choice of $\omega(t)$, then appropriate phase-shift operators are applied before and after the squeezing operator to compensate the extra phase shift.

The condition of $b(-T/2)=b(T/2)=1$ implies that $\omega_-^2 = \omega_0^2$ before and after the operation.  According to Eq.~(\ref{omega-}), this variation of $\omega_-^2(t)$ implies that the varying double well potential would bring two ions to proximity and then separated, as we have expected at the construction.

\section{Construction of beam splitter \label{bs construction}}

Similar to the discussion in Sec. \ref{phase-shift operator}, there are countless forms of $\omega_-(t)$ that can achieve a phase-shift operator on the breathing mode.  It is generally difficult to guess the trapping potential magnitudes, $A(t)$ and $B(t)$, that can produce an appropriate $\omega_-(t)$; they could be chosen only by trial-and-error.
Alternatively, we outline the procedure of acquiring $A(t)$ and $B(t)$ inversely from an intellectually speculated $b(t)$.  

Consider the operation of double well runs from $-T/2$ to $T/2$.  $b(t \leq -T/2)$ and $b(t \geq T/2)$ are the necessary condition for a phase-shift operator.  An example would be 
\begin{equation}\label{exponential b}
 b_{B}(t)=1-k e^{-t^2/\sigma^2}~;
 \end{equation}
where
$\sigma\ll T$ determines the time scale of the operation; the value $k$ is chosen to generate the desired phase shift.  

The corresponding $\omega_-(t)$ can be deduced inversely from Eq.~(\ref{b equation}).  The time variation of the ion separation, $r(t)$, is then obtained from $\omega_-^2(t)$ by Eq.~(\ref{omega-}).  A constraint on $A(t)$ and $B(t)$ is obtained using the equation of motion Eq.~(\ref{EOM}) and the time variation of $r(t)$.   Together with Eq.~(\ref{constant omega+}), the unique form of $A(t)$ and $B(t)$ can then be found.

\section{Accuracy of beam splitter \label{bs accuracy}}

The quality of the beam splitter operation would be degraded by the anharmonic composition of the Hamiltonian $\hat{\mathcal{H}}_B$, which are higher than second order quantum fluctuation that do not covered by the approximated Hamiltonian $\hat{\mathcal{H}}_2$, i.e.
\begin{eqnarray}\label{BS nonlinear}
\hat{\mathcal{H}}_B &=& \hat{\mathcal{H}}_2 - \frac{\sqrt{2} e^2 \hat{q}_-^3}{2\pi \epsilon_0 r^3(r+\sqrt{2}\hat{q}_-)}+\sqrt{2}A(t)r(3 \hat{q}_+^2 \hat{q}_- + \hat{q}_-^3) \nonumber \\
&& +\frac{A(t)}{2} (\hat{q}_+^4 + 6\hat{q}_+^2 \hat{q}_-^2 + \hat{q}_-^4)~.
\end{eqnarray}
When comparing with $\hat{\mathcal{H}}_2$, the magnitude of the anharmonic terms is roughly offset by $\sqrt{\langle \hat{q}^2\rangle}/r$ that scales as $\sqrt{\bar{n}}$, where $\bar{n}$ is the average number of phonon involved in the simulation.

The actual error produced crucially depends on the $b(t)$ chosen.  To demonstrate the feasibility of our beam splitter, we numerically simulate a 50:50 beam splitter using the $b_B(t)$ in Eq.~(\ref{exponential b}).  Only the steps involving the double well are investigated (step II-IV in Fig.~\ref{fig:bs}); the pick-up process and the harmonic well transportation are assumed to be error-free.  The evolution of the motional states is tracked by numerically integrating the Schr\"{o}dinger equation with the Hamiltonian in Eq.~(\ref{BS nonlinear}).  For simplicity, the interaction terms between the + and - modes are replaced by the expectation value, e.g. $\hat{q}_+^2 \rightarrow \langle \hat{q}_+^2 \rangle$, in the evolution of - mode; this treatment is accurate in our case because the back reaction scales as $\sqrt{\langle \hat{q}^2 \rangle}/r$ that is small.  

We consider the ions are $^{40}$Ca$^+$ and the trapping frequency of the storage traps are $\omega_0=2\pi$ MHz.  The pick-up position is set as $50~l_0$ from the mid-point of the ions, where $l_0\equiv \sqrt[3]{e^2/4 \pi \epsilon_0 m \omega_0^2}\approx 4.45~\mu$m is the ions' separation if they are placed in a single harmonic well.  The operation speed is adjusted by tuning the characteristics time $\sigma$, and the value of $k$ is respectively chosen for each $\sigma$ to generate the desired phase shift.  A quality beam splitter should produce a final state whose phase and fidelity are close to those in the ideal case.

\begin{figure}
\centering
\includegraphics{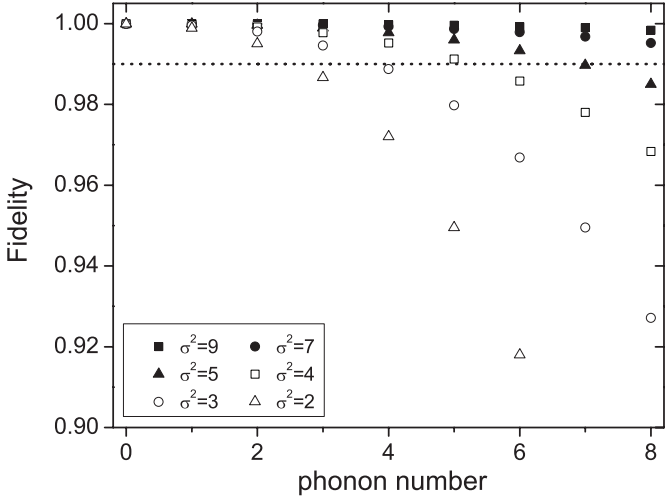}
\caption{\label{fig:bs result} Fidelity of the phonon state after a 50:50 beam splitter operation with $\omega_0^2 \sigma^2 = 2, 3, 4, 5, 7, 9$.  Total time for the double well to bring ions from and back to the pick-up position (step II to IV in Fig.~\ref{fig:bs}) are about $11, 13, 15, 17, 20, 22\times 1/\omega_0$ respectively.  The minimum separation between ions is about 1 $l_0$ in all the six runs.  The dotted line shows a benchmark of 0.99 fidelity.}
\end{figure}

We set the input states of both the centre-of-mass mode and the breathing mode to be pure Fock states, i.e. $|\psi(-T/2)\rangle= |n_+\rangle |n_-\rangle$, such that the final states should be the same as the input states up to a phase.  We find that the output fidelity of any $|n_+\rangle |n_-\rangle$ input is higher than that of a $|\max\{n_+,n_- \} \rangle |\max\{n_+,n_- \} \rangle$ input, so only the runs with $n_+=n_-$ are shown in Fig.~\ref{fig:bs result} for comparison.  The double well process generates less than $1\%$ error when $\omega_0^2\sigma^2 \gtrsim 7$ for $n_{\textrm{max}} \leq 8$; the total process time is about only $20/\omega_0 \approx 3.1~\mu$s.  In addition, the phase errors are less than $1\%$ in all the runs.  We conclude that for a bosonic simulation with single digit number of phonons in each mode, a quality phonon beam splitter can be implemented within a few $\mu$s. 

The accuracy of the beam splitter is worsened, as expected, when more phonons are involved, because the factor $\sqrt{\langle \hat{q}^2 \rangle}/r$ increases.  A higher operation speed also exacerbates the error.  When considering Eq.~(\ref{BS nonlinear}) in the interaction picture with respect to $\hat{\mathcal{H}}_2$, the terms that are third order to $\sqrt{\langle \hat{q}^2\rangle}/r$ are off-resonant.  These nonlinear terms' contributions are suppressed by the RWA if the terms' variation is slow, then the effective Hamiltonian would be reduced to the forth order of $\sqrt{\langle \hat{q}^2\rangle}/r$.  On the other hand, the off-resonant terms are significant for a high speed operation because the RWA is less effective.

\section{Ion separation without heating \label{ion separation}}

Usual processes of ion separation is to first reduce the trapping potential, then an additional potential is raised in the middle of the trap \cite{Rowe:2002p6849}.  A fast process would heat up the ion significantly.  In quantum computation, the side effect of heating can be compensated by subsequent cooling that do not disturb the quantum information encoded in the electronic states.  However, the time taken by ion cooling limits the operation speed of the ion trap quantum computer.  Here we propose an alternative method to separate the ions, which can be rapidly conducted with minimal heating, by using an extra quartic potential.

The configuration we are considering is the same as that in Sec. \ref{phonon beam splitter}, where the Hamiltonian of the system is given in Eq.~(\ref{BS H1}).  The ions are initially in the motional ground state at a common harmonic well with trapping frequency $\omega_0$, i.e. $A(0)=0$ and $B(0)=m\omega_0^2/2$.  Our aim is to separate the ions to distant storage traps with frequency $\omega_0$ with no motional excitation apart from the errors caused by the nonlinear terms.  The initial centre-of-mass mode frequency is $\omega_0$ and the breathing mode frequency is $\sqrt{3}\omega_0$ \cite{James:1998p2256}.  Similar to the phonon beam splitter, we require $3 A(t)r^2/m + 2B(t)/m=\omega_0^2$ such that there is no excitation in the centre-of-mass mode.  For the breathing mode, it can be shown that $b(t\leq 0)=1/\sqrt[4]{3}$ is equivalent to the case of a single harmonic trap.  Any $b(t)$ with the boundary conditions
\begin{equation}
b(t\leq 0)=\frac{1}{\sqrt[4]{3}}~,~b(t\geq T)=1,
\end{equation}
can give the potential variation that retain the breathing mode in the ground state.  $\dot{b}(t)$ and $\ddot{b}(t)$ should also be continuous throughout the process unless $r$ changes extremely fast.  An example is 
\begin{equation}\label{separate b}
b_E(t) = \left(\frac{1}{\sqrt[4]{3}}-1\right) e^{-t^3/\sigma^3}+1~,
\end{equation}
where $\sigma$ is some characteristic time scale of the operation.  

\begin{figure}
\centering 
\includegraphics{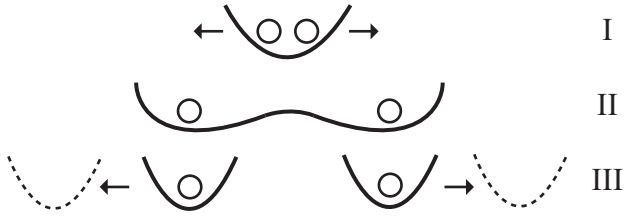}
\caption{\label{fig:separate} Variations of potentials during heatingless ion separation. Step I, a quartic potential is added to the common trap to form a double well.  Step II, double well extends and separates the ions to pick-up position.  Step III, harmonic wells pick-up the ions and bring them to storage traps.}
\end{figure}

After the ions' separation is large enough that the Coulomb interaction is negligible, the double well potential can be switched off and then the ions are picked up by two moving harmonic well.  The velocities of the ions at the pick-up locations can be calculated by time-differentiating Eq.~(\ref{omega-}) once with the $\omega_-^2(t)$ obtained from Eq.~(\ref{b equation}).  The two displacing harmonic wells, which the trapping frequencies are $\omega_0$, should decelerate the ion motion to rest at the storage traps; the time variation of the potential can be deduced from inverse engineering method in Ref. \cite{Torrontegui:2011p9810}.  The whole process is summarized in Fig.~\ref{fig:separate}. Inverting the whole ion separation process can bring two individually trapped ions to a common trap.  

Similar to the phonon beam splitter, the error of this ion separation method is produced by the anharmonicity of the quartic potential and the Coulomb interaction, as shown in Eq.~(\ref{BS nonlinear}).  The magnitude of the generated errors highly depends on the $b(t)$ chosen, but a faster ion separation process would generally create more error because the RWA becomes less effective.  

To demonstrate the feasibility of our scheme, the evolution equation Eq.~(\ref{BS nonlinear}) is numerically integrated using the $b_E(t)$ in Eq.~(\ref{separate b}) with $\sigma=2$.  For simplicity, we assume the operators of the other mode can be approximated by the expectation value, i.e. $\hat{q}^n_\pm\approx \langle \hat{q}^n_\pm \rangle$ in the equation of the $\mp$ mode.  This assumption is accurate because the back reaction only produces higher order corrections to the to the already small errors.  Our calculation simulates only the process of the double well extension (step I to II in Fig.~\ref{fig:separate}; heating effects caused by the pick-up and harmonic well transportation are negligible if the processes are well controlled.  We consider the ions are Ca$^+$ and $\omega_0=2\pi \times 10^6$ Hz, where the initial separation $r_0\approx 4.45 ~\mu$m.  After separating the ions from $r_0$ to about $100~r_0$, only about 0.001 quanta is excited for both the + mode and the - mode.  The duration of this process is about $5\times 1/\omega_0\approx 0.8 \mu$s.  Because the transportation of the ions from the pick-up position to the storage traps (step III in Fig.~\ref{fig:separate}) can be arbitrarily fast \cite{Torrontegui:2011p9810}, the time duration of the whole ion separation process can be as short as 1 $\mu$s with minimal heating.

\bibliographystyle{phaip}
\pagestyle{plain}
\bibliography{boson2}

\end{document}